\documentclass[%
reprint,
superscriptaddress,
showpacs,preprintnumbers,
amsmath,amssymb,
aps,
prl,
]{revtex4-2}
\usepackage{color}
\usepackage{graphicx}
\usepackage{bm}
\usepackage{mathrsfs}
\usepackage{dsfont}
\usepackage{bbm}
\usepackage[colorlinks,allcolors=blue]{hyperref}

\makeatletter

\newcommand{\Rmnum}[1]{\expandafter\@slowromancap\romannumeral #1@}

\makeatother

\begin{document}
\title{Retrieving High-Dimensional Quantum Steering From a Noisy Environment with $N$ Measurement Settings}
	
\author{Rui Qu}
	\affiliation{MOE Key Laboratory for Nonequilibrium Synthesis and Modulation of Condensed Matter, School of Physics, Xi’an Jiaotong University, Xi'an 710049, China}
	
	\author{Yunlong Wang}
	%\email{yunlong.wang@mail.xjtu.edu.cn}
	\affiliation{MOE Key Laboratory for Nonequilibrium Synthesis and Modulation of Condensed Matter, School of Physics, Xi’an Jiaotong University, Xi'an 710049, China}
	\affiliation{Shaanxi Key Laboratory of Quantum Information and Quantum Optoelectronic Devices, School of Physics, Xi'an Jiaotong University, Xi'an 710049, China}
	
	\author{Min An}
	\affiliation{MOE Key Laboratory for Nonequilibrium Synthesis and Modulation of Condensed Matter, School of Physics, Xi’an Jiaotong University, Xi'an 710049, China}
	
	\author{Feiran Wang}
	\affiliation{School of Science, Xi’an Polytechnic University, Xi'an 710048, China}
	
	\author{Quan Quan}
	\affiliation{MOE Key Laboratory for Nonequilibrium Synthesis and Modulation of Condensed Matter, School of Physics, Xi’an Jiaotong University, Xi'an 710049, China}
	\affiliation{Shaanxi Key Laboratory of Quantum Information and Quantum Optoelectronic Devices, School of Physics, Xi'an Jiaotong University, Xi'an 710049, China}
	
	\author{Hongrong Li}
	\affiliation{MOE Key Laboratory for Nonequilibrium Synthesis and Modulation of Condensed Matter, School of Physics, Xi’an Jiaotong University, Xi'an 710049, China}
	\affiliation{Shaanxi Key Laboratory of Quantum Information and Quantum Optoelectronic Devices, School of Physics, Xi'an Jiaotong University, Xi'an 710049, China}
	
	\author{Hong Gao}
	\affiliation{MOE Key Laboratory for Nonequilibrium Synthesis and Modulation of Condensed Matter, School of Physics, Xi’an Jiaotong University, Xi'an 710049, China}
	\affiliation{Shaanxi Key Laboratory of Quantum Information and Quantum Optoelectronic Devices, School of Physics, Xi'an Jiaotong University, Xi'an 710049, China}
	
	\author{Fuli Li}
	\affiliation{MOE Key Laboratory for Nonequilibrium Synthesis and Modulation of Condensed Matter, School of Physics, Xi’an Jiaotong University, Xi'an 710049, China}
	\affiliation{Shaanxi Key Laboratory of Quantum Information and Quantum Optoelectronic Devices, School of Physics, Xi'an Jiaotong University, Xi'an 710049, China}
		
	\author{Pei Zhang}
	\email{zhangpei@mail.ustc.edu.cn}
	\affiliation{MOE Key Laboratory for Nonequilibrium Synthesis and Modulation of Condensed Matter, School of Physics, Xi’an Jiaotong University, Xi'an 710049, China}
	\affiliation{Shaanxi Key Laboratory of Quantum Information and Quantum Optoelectronic Devices, School of Physics, Xi'an Jiaotong University, Xi'an 710049, China}
	\date{\today}
	
\begin{abstract}
	One of the most often implied benefits of high-dimensional (HD) quantum systems is to lead to stronger forms of correlations, featuring  increased robustness to noise.
	Here, we experimentally demonstrate the $n$-setting linear HD quantum steering criterion. We verify the large violation of the steering inequalities without full-state tomography. The lower bound of the violation is $2.24\pm0.01$ in 11 dimensions, exceeding the bound ($V<2$) of 2-setting criteria. Hence, a higher strength of steering has been revealed. Moreover, we demonstrate the method for enhancing the noise robustness without increasing dimension, alternatively, by increasing measurement settings. Using the entanglement in 11 dimensions, we experimentally retrieve steering nonlocality with $63.4\pm1.4\%$ isotropic noise fraction, surpassing the $50\%$ limitation of 2-setting criteria. Our work offers the potential for practical one-sided device-independent quantum information processing that tolerates the noisy environment, lossy detection, and transcends the present transmission distance limitation.
\end{abstract}
	
	\maketitle  
	
	%\label{sec:introduction}
	%%%%%%%%%%%%%%%%%%%%%%%%%%%%%%%%
	\textit{Introduction.---}One of the most fundamental characteristics of quantum systems is that distant objects may show stronger correlations than in the classical world. Quantum correlations can be strictly categorized into three hierarchies  \cite{wiseman2007prl,quintino2015inequivalence}: Entanglement, Einstein-Podolsky-Rosen~(EPR) steering and  Bell nonlocality, among which entanglement is the weakest and Bell nonlocality is the strongest. As a generalization of EPR paradox \cite{einstein1935can}, steering was first introduced by Schrödinger in 1935 to describe Alice's ability to remotely affect Bob's state via local measurements \cite{schrodinger1935}. 
	The modern notion and operational framework of steering were formulated in the innovative work of Wiseman $et\ al.$ \cite{wiseman2007prl} where they defined steering in terms of the impossibility to describe the conditional states at one party by the local hidden state (LHS) model. The certification of quantum steering is an asymmetric quantum information task where one party's measurement devices are untrusted \cite{uola20RMP-review,16review-steering-cavalcanti2016quantum,09review-steering-reid2009colloquium}. This makes steering have many applications, such as subchannel discrimination \cite{subchannel-watrous2015prl,subchannel-sun2018npj}, one-sided device-independent (1SDI) quantum key distribution \cite{Wiseman12pra-SDI-QKD}, secure quantum teleportation \cite{tele-he2015prl,tele-reid2013pra} and 1SDI randomness generation \cite{random-law2014quantum,skrzypczyk2018prlmaximal,guo2019prl-random,wang2018science}. 
		
	High-dimensional (HD) quantum system is expected to transcend limitations of qubits in some applications \cite{cozzolino2019-HDreview}, due to its stronger nonlocal correlation \cite{zeilinger02prl,thew2004prl-bell}, larger channel capacity \cite{zeilinger2006-capacity,hu2018beating,islam2017sa-provably} and higher robustness against noise \cite{marcus19prx-noise,vertesi2010prl-noise,zhu19arxiv-noise}. 
	Recently, some works began to explore the HD steering effect \cite{designolle2020genuine,zeng2018prl-HDsteering,wang2018science,li2015PRLgenuine,guo2019prl-random}. In particular, Zeng $et\ al.$ exhibited the strong noise robustness increasing with the extra dimension in HD steering \cite{zeng2018prl-HDsteering}. 
	However, these works are restricted to the 2-setting steering criteria where the degree of violation is lower than 2, which means the revealed strength of steering is low. This demands the noise threshold of HD entangled system for exhibiting steering to be below $50\%$ even if the dimension $d$ is infinity, otherwise the steerability cannot be detected. Hence, these methods would limit the highly efficient and  noise-robust detection of HD steering. 
	
	In order to solve these problems, we put the HD  steering tests in a condition with multiple measurement settings. Based on a complete set of mutually unbiased bases (MUBs), one can construct the $n=(d+1)$-setting steering criterion in dimension $d$, showing an unbounded violation of the steering inequality \cite{marciniak2015prl-unbounded,zhu16prl-renyi}. Interestingly, there is no similar simple form in the existing Bell inequalities with large violation \cite{junge2010unbounded,junge2011large,bell-06PRA,yin15PRA-bellunbounded}. As the degree of nonlocality depends on the strength of quantum steering and uncertainty relationship \cite{wehner2010uncertainty}, this unbounded violation indicates particular properties of steering. Compared with the 2-setting criteria, the extra measurement settings reveal more information about the underlying quantum state, thus show a higher strength of steering. This would feature a higher noise threshold \cite{skrzypczyk2015pra-loss}. As the dimension tends to infinity, one can overcome unbounded amount of noise to detect steerability with $n$ settings. The multisetting steering criteria would be of fundamental interest and provide practical applications in the 1SDI quantum information processing, whose demonstration is still lacking.
	
	Here we implement an experiment based on photons entangled in their orbital angular momentum (OAM) to explore HD steering effect with $n$ settings.  
	We first experimentally demonstrate the large violation of steering inequalities using measurements onto $d+1$ MUBs. The results show a bound of the violation $V\ge2.24\pm0.01$ in dimension $d=11$, while that in the 2-setting criteria is limited by $V<2$. In other words, a higher strength of steering has been revealed. Then, by introducing a tunable isotropic noise into the prepared  11-dimensional maximally entanglement, we experimentally retrieve steering with $63.4\pm1.4\%$ noise fraction. In dimensions $d=5,7$, the experimental noise thresholds also exceed $50\%$, while $50\%$ is the limitation of 2-setting criteria whatever the dimension is. Finally, we demonstrate the method for enhancing the noise robustness without increasing dimension, alternatively, by increasing measurement settings. Our work makes a big step towards practical 1SDI quantum information processing that can tolerate the noisy environment, lossy detection and long-distance transmission.
	
	%\section{Theory} % (fold)
	%\label{sec:Unbounded Violation of Steering Inequalities}
	\textit{Theory.---}Consider a scenario featuring in two space-like separate observers, Alice and Bob, sharing a bipartite state $\rho_{AB}$. Alice chooses one of her measurement settings $\left\{A_{a|x}\right\}_x$, labeled by $x=1,...,m$ and receives a result $\ a \in \left\{\ 1,...,d\right\}$, where $A_{a|x} $ denotes the positive operators satisfying $\sum\nolimits_a {{A_{a|x}}}=\mathbbm{1}_A$ for each $x$, thus remotely steers Bob's subsystem to the conditional state
	\begin{equation}
	{\sigma _{a|x}} = {\rm{T}}{{\rm{r}}_A}\left[ {\left( {{A_{a|x}} \otimes \mathbbm{1}_B} \right){\rho _{AB}}} \right].
	\end{equation}
	Bob performs his measurements on the conditional state and confirms whether the correlation of their results violates the steering inequalities.
	
	Here we consider the scenario where Alice and Bob share a HD maximally entanglement  $\left| \Phi _d \right>  = 1/\sqrt d \sum\nolimits_{l = 0}^{d - 1} {\left| {l,l} \right> }$ and perform ideal measurements on the complete set of MUBs. Two orthonormal bases are called MUBs if the inner product of any vector from the first basis with any vector from the second basis is $\left|\left< {\varphi _x^a} | {\varphi _y^b} \right>\right| = 1/\sqrt d $ \cite{bengtsson2007-mub}. When the dimension $d$ is a prime power, we can always construct $n=d+1$ MUBs using the standard construction \cite{wootters1989-mub}. Then we introduce the linear steering functional, which is expressed as \cite{marciniak2015prl-unbounded}
	\begin{align}\label{eq2}
	\ S_{\rm{Q}} &= \sum_{x=1}^{n}\sum_{a=1}^{d} {\rm{Tr}}\left[ {\left( {{A_{a|x}} \otimes A_{a|x}^T} \right)\rho _{AB} } \right]\nonumber \\ 
	&=\sum_{x=1}^{n}\sum_{a=1}^{d}P\left(a,a|x,x\right),
	\end{align}
	where $P\left(a,a|x,x\right)$ denotes the joint probability that both parties obtain result $a$ when measuring in the same basis $x$. Regardless of noise and loss, the maximal value of Eq.(\ref{eq2}), called the quantum bound, equals $S_{\rm{Q}}=n$.
	
	If they share a separable state $\rho = \sum\nolimits_i {{p_i}\rho _i^A \otimes\rho _i^B} \ $, Alice can only steer Bob's subsystem to the conditional state that satisfies the LHS model
	\begin{align}
		\sigma _{{a|x}}^{{\rm{LHS}}} = \sum_i {{p_i}{\rm{T}}{{\rm{r}}_A}\left[ {{A_{a|x}}\rho _i^A} \right] \otimes \rho _i^B=\sum_{i}q(a|x,i)\rho_i^B}.
	\end{align}
	The maximum value of the linear steering functional for the LHS model is bounded by \cite{skrzypczyk2015pra-loss}
	\begin{align}
		\ {S_{{\rm{LHS}}}} = \underset{\sigma_{a|x}^{\rm{LHS}}}{\rm{max}}\sum_{x=1}^{n} {\sum_{a=1}^{d} {{\rm{Tr(}}{A_{a|x}}\sigma _{{a|x}}^{{\rm{LHS}}})} }  \le 1 + \frac{n-1}{\sqrt d }.
	\end{align}
	In particular, ${S_{{\rm{LHS}}}}$ equals to the boundary for $n=2$, that is, ${S_{{\rm{LHS}}}}=1+1/\sqrt{d}$~\cite{zeng2018prl-HDsteering}. If Bob demonstrates the violation of the inequality $S_{\rm{Q}}\le1+(n-1)/\sqrt d$, he has to admit they share steerable states. The degree of violation, a natural quantifier of steering, can be defined as $V=S_{\rm{Q}}/S_{\rm{LHS}}$. Thus, its lower bound is obtained, 
	\begin{align}\label{eq5}
	V=\frac{S_{\rm{Q}}}{S_{\rm{LHS}}}\ge\frac{n}{1 + (n-1)/\sqrt{d}} .
	\end{align}
	Fixing $n=d+1$, we can acquire an unbounded violation of order $O(\sqrt d)$, while the violation in 2-setting criteria is limited by $V=2/({1+1/\sqrt{d}})<2$. 
	\begin{figure}[!b]
		\centering
		\includegraphics[width=0.65\linewidth]{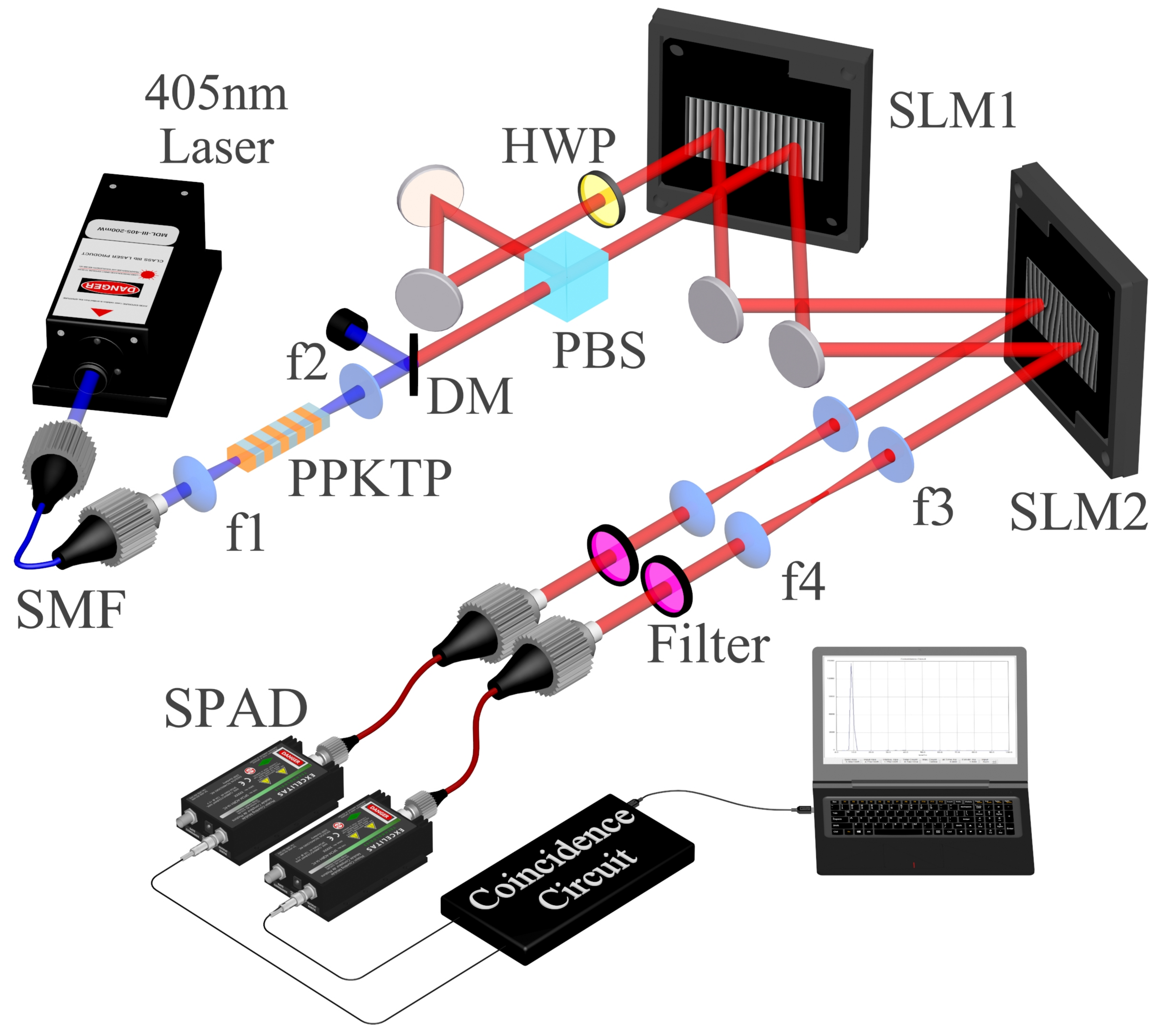}
		\caption{Experimental setup. Pumper: diode laser centered at 405~nm; SMF: single mode fiber; PPKTP: periodically poled potassium titanyl phosphate; $\rm{f}_{1,2,3,4}$: lens; DM: dichroic mirror; PBS: polarizing beam splitter; HWP: half-wave plate; SLM: spatial light modulator; Filter: high-pass optical filter; SPAD: single-photon avalanche detector.}\label{setup}
	\end{figure}
	\begin{figure*}[!t]
		\centering
		\includegraphics[width=0.9\linewidth]{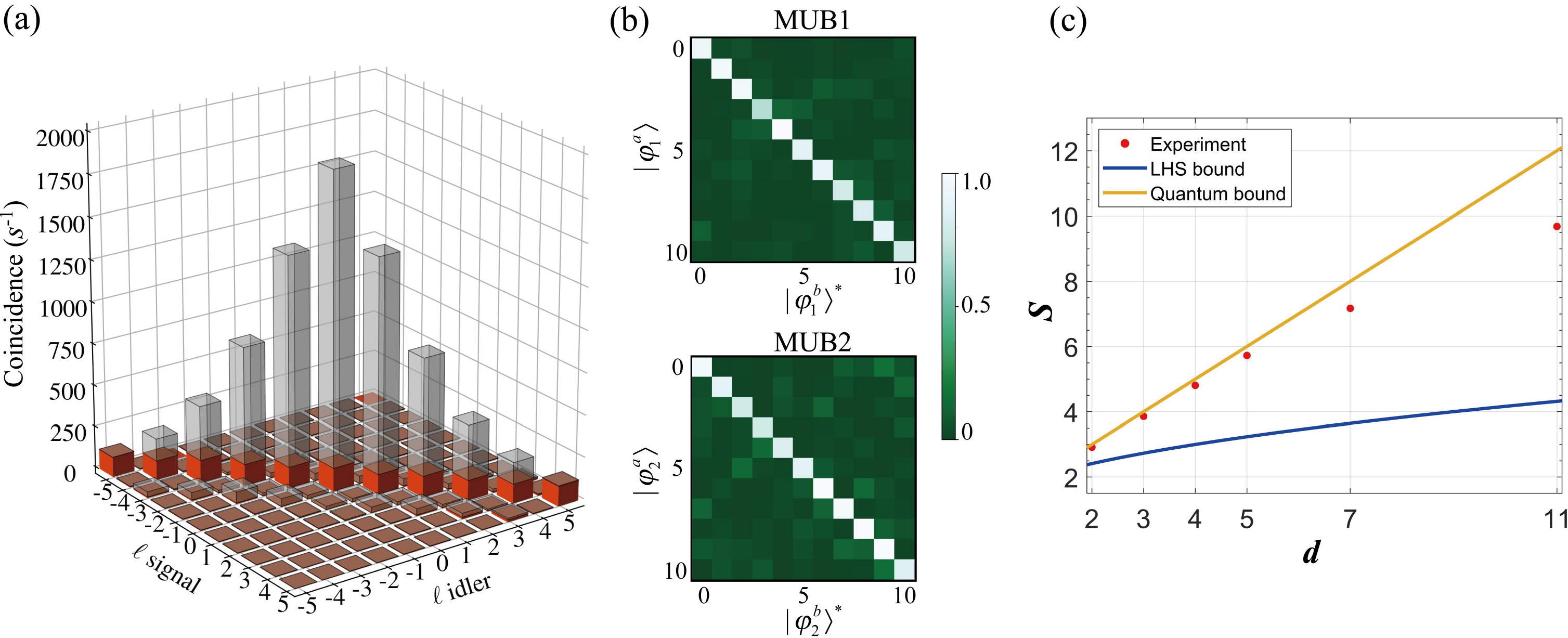}
		\caption{Experimental results. (a) Experimental 2-photons coincidence counts in computational basis $\left|l_{i}\right>_A\left|l_{j}\right>_B$ with $l_{i},l_{j}=-5,...,5$. The coloured and transparent bars depict the coincidence counts with and without entanglement concentration. (b) Normalized 2-photons coincidence rates in two 11-dimensional MUBs ($x = 1$ and $2$). When Alice's projective measurement is $\left|\varphi_x^a\right>$, Bob measures on the complex conjugation bases ${\left|\varphi_x^b\right>}^*$($a,b=0,...,10$). All the actual coincidence rates in 11 dimensions are in Supplemental Material. (c) Experimental results of the steering functional $S$  with dimension up to 11. The error bars are of the order of $10^{-2}$, much smaller than the marker size.}\label{fig:2}
	\end{figure*}
	
	This large violation leads to a higher noise threshold in the steering test. Consider a $d$-dimensional isotropic state $\rho _{\rm{iso}}=p\left|\Phi_d\right>\left<\Phi_d \right|+(1 - p)\mathbbm{1}/d^2 $, where $p$ denotes the  mixing parameter of the pure state. 
	The steering functional for the isotropic state can be calculated by $S_{\mathrm{iso}}=n[p+(1-p)/d]$.	To violate the steering inequality, the lower bound of the mixing parameter is demanded:
	\begin{align}
		p_{\mathrm{min}}(d,n)=\frac{(n+\sqrt{d})(\sqrt{d}-1)}{n(d-1)}.
	\end{align}
	For $n=d+1$, this value is ${p^{(n)}_\mathrm{min}}=(d^{3/2}-1)/(d^2-1)$ while that in the linear 2-setting criteria is ${p^{(2)}_\mathrm{min}}=1/2[1+(\sqrt{d}-1)/(d-1)]$.  
	Importantly, as the dimension tends to infinite, $ p_\mathrm{min}^{(n)}$ tends to 0, while ${p^{(2)}_\mathrm{min}}$ tends to $50\% $.
		
	\textit{Experimental demonstration of large violation.---}The setup is shown in Fig.~\ref{setup}. A CW 405~nm laser beam is first coupled to a single-mode fiber (SMF) to be spatially filtered, and then focused by a 250~mm lens $f_1$ to a spot size of 0.5~mm at a 5-mm-long nonlinear PPKTP crystal. The type-II spontaneous parametric down-conversion (SPDC) process generates orthogonally polarized pairs of OAM entangled photons at 810~nm. Then, the photon pairs are recollimated by a 100~mm lens $f_2$, and separated by a polarization beam splitter (PBS).
	A half-wave plate (HWP) is used to rotate the polarization of the idler photon from vertical to horizontal, allowing it to be manipulated by the spatial light modulator (SLM). We make the idler photon reflect odd times to reverse its OAM from $-l$ to $l$ and thus obtain the original HD entanglement $\left| \Phi \right> = \sum\nolimits_{ - \infty }^\infty {{c_l}{{\left| l \right> } _A}{\left| l \right> }_B} $, where $l$ is the index of OAM. The SLM1 is loaded with a regular diffraction grating. Projective measurements on the OAM basis and its MUBs are performed by the SLM2 loaded with computer-generated holograms (CGHs), a group of so called intensity-flattening lenses~\cite{18OE-ift} and SMFs. The joint probability distributions are obtained by measuring coincidence counts using single-photon avalanche detectors (SPADs) connected to a coincidence circuit, and then used to evaluate the steering functional in Eq.~(\ref{eq2}).
	
	The original OAM entanglement produced by SPDC process is not a maximally entangled state owing to the limited spiral bandwidth \cite{torres03PRAbandwidth}. To faithfully demonstrate the large violation of steering inequalities, we use the entanglement concentration technique called Procrustean method \cite{bennett1996concentrating} to equalize different orders of OAM. This is generally done by choosing local operations matched to the spiral bandwidth of our SPDC source \cite{dada2011nphys}. Fig.~\ref{fig:2}(a) shows our coincidence results before and after entanglement concentration. 	

	To test within $d$-dimensional subspace, for odd $d$, we choose the modes $l$ from $–(d-1)/2$ to $(d-1)/2$ as the computational basis. For even $d$, we choose $l$ from $-d/2$ to $d/2$, omitting the $l=0$ mode. And we construct the other $d$ MUBs using these modes, respectively. The normalized coincidence rates in two 11-dimensional MUBs ($x = 1$ and $2$) are shown in Fig.~\ref{fig:2}(b).
	\begin{figure}[!b]
		\centering
		\includegraphics[width=0.8\linewidth]{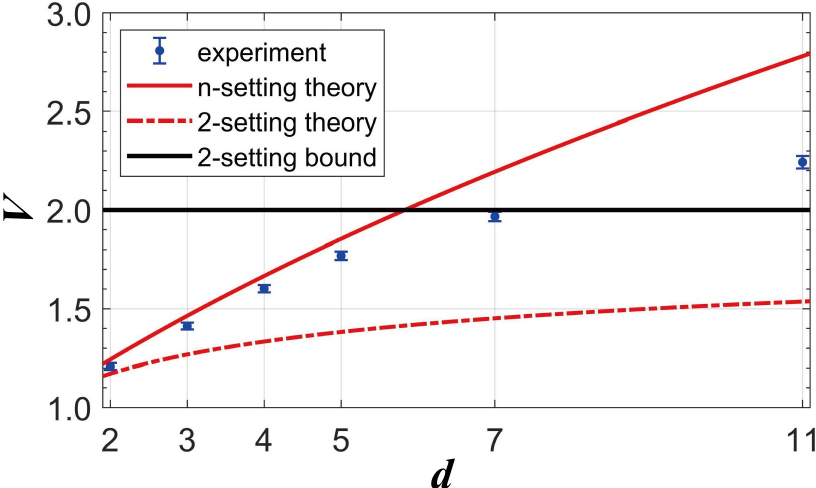}
		\caption{Experimental results of the bounds of the violation $V$ in the $n$-setting steering criteria with dimension up to 11. The error bars represent 3 standard deviations.}\label{fig:3}
	\end{figure}
    \begin{figure*}[!t]
	\centering
	\includegraphics[width=0.8\linewidth]{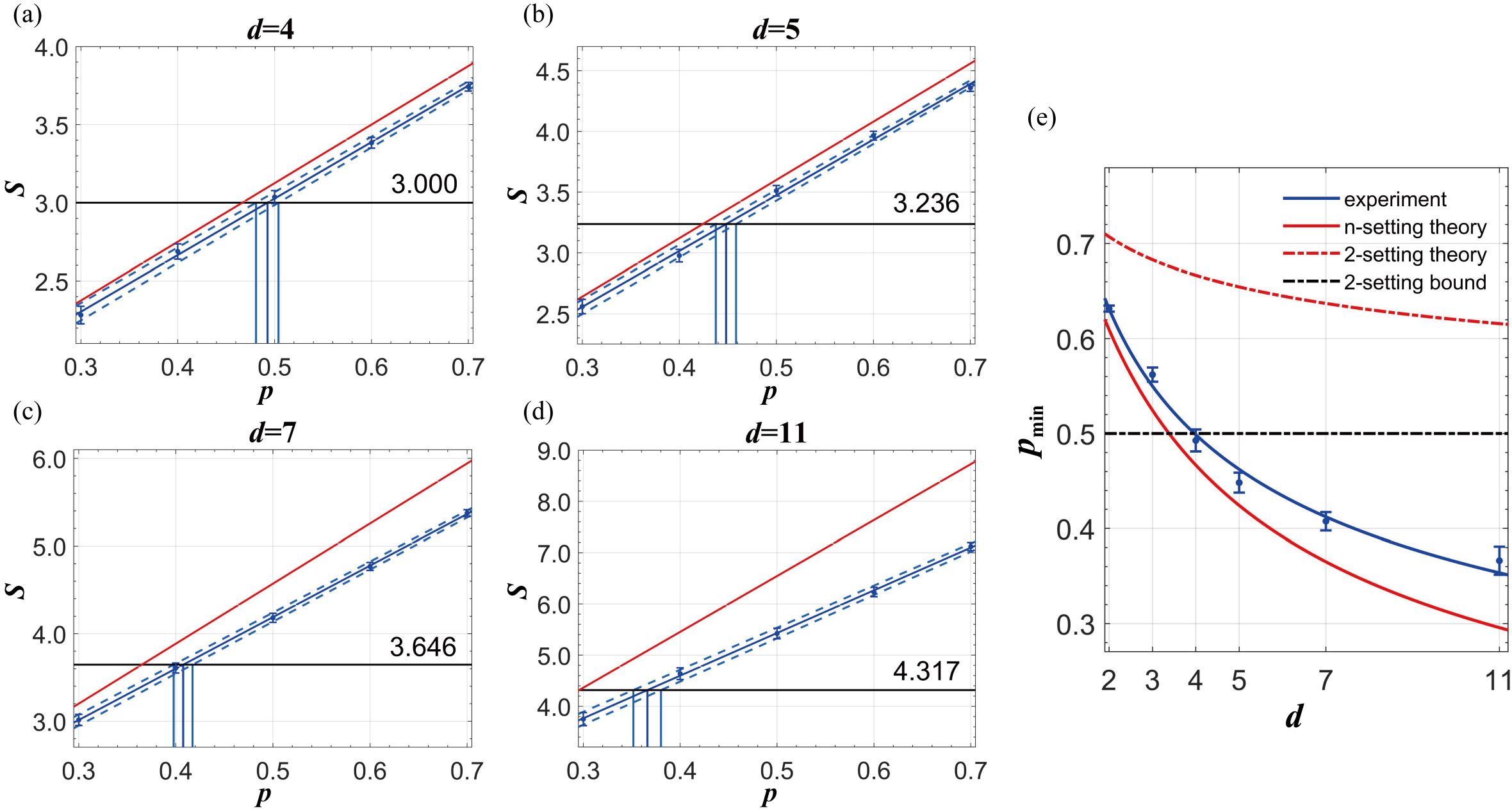}
	\caption{Experimental results of the noise threshold. (a)-(d) The steering functional of isotropic state $S_{\rm{iso}}$ for different mixing parameters in dimensions of $d=4,5,7,11$. The blue inclined lines are fitted by the blue experimental points, and the blue dashed lines represent the range of error bars. The red lines are the theoretical prediction for the isotropic states $S_{\mathrm{iso}}=n[p+(1-p)/d]$, and the black horizontal lines represent the bounds of the LHS steering functional. The blue vertical lines denote the experimental lower bounds of the mixing parameter and their errors. (e) The theoretical and experimental values of the lower bound on mixing parameter $\ p_{\mathrm{min}}$ as a function of dimension $d$. The theoretical prediction of the 2-setting criterion and its asymptotes are given for comparison.}\label{fig:4}
    \end{figure*}
	
	Fig.~\ref{fig:2}(c) displays our experimental results of steering functional. The quantum bound and LHS model bound are presented for comparison. As we can see, all the values of $S_{\rm{Q}}$ exceed the corresponding LHS bounds, which means we certify HD quantum steering with dimension up to $d=11$. 

	Then, the experimental degrees of violation in different dimensions are obtained by $V=S_{\rm{Q}}/S_{\rm{LHS}}$ and displayed in Fig.~\ref{fig:3}. As we can see, all the experimental lower bounds exceed the theoretical values of the 2-setting criterion. In particular, the experimental violation is $V\ge2.24\pm0.01$ in dimension $d=11$, exceeding the upper bound of 2-setting criteria. It means our results reveal a high strength of steering which cannot be realized using 2-setting criteria.
	
	% section experiment1 (end)
		
	%\label{sec:experiment 2}	

    \textit{Experimental verification of the noise robustness in $n$-setting HD steering.---}To experimentally prepare the isotropic state $\rho _{\rm{iso}}=p\left|\Phi_d\right>\left<\Phi_d \right|+(1 - p)\mathbbm{1}/d^2 $, we follow Ref.~\cite{zeng2018prl-HDsteering} to construct a simplified  isotropic noise in a statistical sense:   
    \begin{align}\label{iso}
    \mathbbm{1}/d^2=&\frac{1}{(d+1)d^2}[\sum_{i,j=0}^{d-1}{|l_i l_j\rangle}_{\mathrm{AB}}\left\langle l_{i}l_j\right|\nonumber\\
    &+\sum_{x=0}^{d-1}\sum_{a,b=0}^{d-1}{|\varphi_x^a \varphi_x^b\rangle}_{\mathrm AB}\left< \varphi_x^a\varphi_x^b\right|],
    \end{align}
	where $\left\{\left|l_i\right>\right\}$ and $\left\{\left| {\varphi _x^a} \right>\right\}_x$ denotes OAM and OAM MUB modes, respectively.
	According to Eq.~(\ref{iso}), $\mathbbm{1}/d^2$ contains $(d+1)d^2$ different two-qudit product states, corresponding to $(d+1)d^2$ CGHs, each of which represents the product state  $\left|l_i\right>_A\left|l_j\right>_B $ or $\left\{\ {\left| \varphi _x^a \right>}_A{{\left| \varphi _x^b \right>}_B}\right\}_x $. When the incident photons are affected by these CGHs, the initial states transform into the corresponding states and covered all elements in Eq.~(\ref{iso}).
	
	We put one CGH of regular diffraction grating which represents the pure state $\left|\Phi_d\right>$ and $(d+1)d^2$ CGHs which represent the isotropic noise into a random pool. By using a random number generator, we control the emerging probability of each CGH, and sample from the random pool to construct a targeted animation displayed in SLM1. Then we calculate the emerging probability for each of these $[1+(d+1)d^2]$ CGHs in the animation with a certain mixing parameter $p$. The isotropic state contains two components $\left|\Phi_d\right>\left<\Phi_d\right|$ and $\mathbbm{1}/d^2$ with probability $p$ and $(1-p)$, respectively. Going through all possibilities of the pure state takes at least $dt$ time, while the ergodic time is $d^2t$ for the isotropic noise. Considering the mixing parameter $p$, the ratio of pure entanglement and mixed isotropic noise is $pdt: (1-p)d^2t$. So the emerging probability of the CGH of regular diffraction grating is $P=p/[p+(1-p)d]$, the other $(d+1)d^2$ CGHs share the probability of $(1-P)/[(d+1)d^2]$. In the experiment, the exposure time of each CGH is set as $t = 0.02$ s. And we set the number of sampling times as $N=500$ to reduce the deviation induced by the sampling randomness. In this manner, we can obtain the tunable isotropic state $\rho _{\rm{iso}}$ by setting a certain $p$.

	The experimental results of $S_{\rm{iso}} $ as a  function of mixing parameter $p$ in dimensions $d=4,5,7,11$ are shown in Fig.~\ref{fig:4}(a), (b), (c) and (d), respectively. We set 5 levels of mixing parameter  $p=0.3$$\sim$$0.7$ as points to fit the blue solid inclined lines, while the blue dashed lines exhibit the range of error bars. Note that the theoretical (red) and experimental (blue) lines are both linearly changed with $p$. Nevertheless, the discrepancies of their slopes becomes larger as dimension increases because of the increasing shot noise and crosstalk of different OAM modes. The black horizontal lines with the corresponding values denote the bound of the LHS steering functional. 
	
	The experimental results of the lower bound on  mixing parameter $p_\mathrm{min}$ in Fig.~\ref{fig:4}(e) are obtained from the points where the blue solid inclined lines intersect the black horizontal lines in Fig.~\ref{fig:4}(a)-(d). As we can see, all the experimental values of the lower bound on mixing parameter are lower than that of 2-setting theory. In particular, the experimental value is $p_\mathrm{min}^{(n)}=36.6\pm1.4\%$  for $d=11$, which means we retrieve steering with $63.4\pm1.4\%$ isotropic noise. For $d=5,7$, the noise thresholds $1-p_\mathrm{min}^{(n)}$ also exceed $50\%$, surpassing the upper bound of 2-setting criteria.
	\begin{figure}[t]
		\centering
		\includegraphics[width=0.8\linewidth]{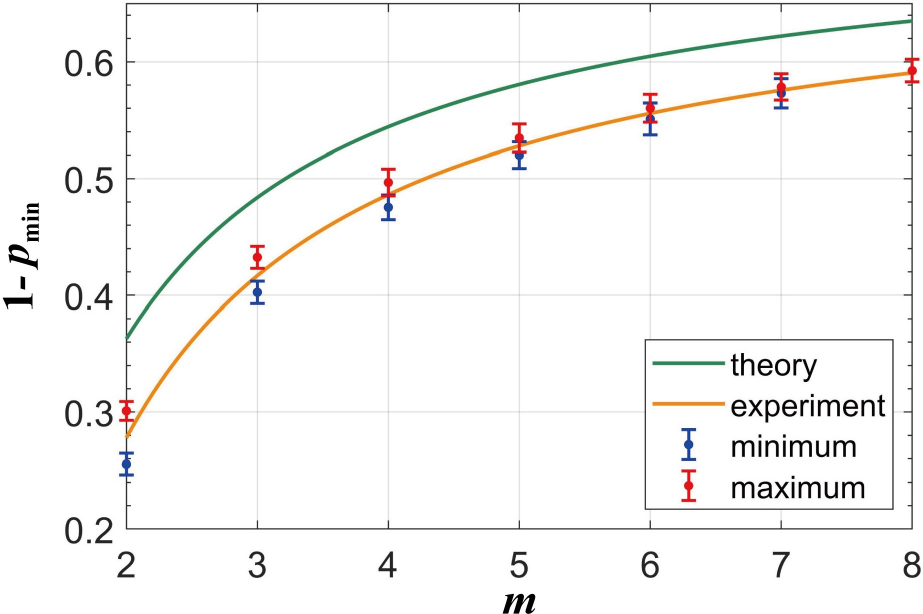}
		\caption{Experimental results of the noise threshold $1-p_\mathrm{min}$ for different numbers of measurement settings $m$ in dimension $d=7$. For each $m$, the minimum and maximum experimental values are provided.}\label{fig:5}
	\end{figure}
	
	Finally, we investigate the method to enhance the noise robustness without increasing dimension, alternatively, by increasing measurement settings. Fixing $d=7$, we experimentally obtain the noise threshold $1-p_\mathrm{min}^{(m)}$ as a function of the number of measurement settings $m$, as shown in Fig.~\ref{fig:5}. Note that choosing $m$ MUBs from the complete set of MUBs, there are $C_{d+1}^m$ possible combinations, giving rise to	potentially different noise thresholds. We provide the minimum and maximum values for simplicity. Although there is a discrepancy between experimental and theoretic curves, the noise threshold increases evidently with increasing measurement settings. Hence, the noise robustness of HD steering effect is indeed enhanced by increasing measurement settings. Moreover, the multisetting method allows one to optimize the number of measurement settings according to the actual amount of noise, leading to the optimal cost of the data acquisition time for certifying HD steering.

	% section experiment 2 (end)
	
	%\section{Discussion} % (fold)
	%\label{sec:Conclusions}
	\textit{Discussion.—}We have experimentally demonstrated HD steering with dimension up to $d=11$ based on the OAM entangled states using $n$ measurement settings.
	The experimental results showed that our method obtained a large violation of the steering inequalities, revealing a high strength of steering. We also demonstrated this would lead to higher noise thresholds for certifying HD steering. We experimentally retrieved steering nonlocality in dimensions $d=5,7,11$ with the noise fractions surpassing the $50\%$ limitation of the 2-setting criteria. These criteria directly uses probability distributions from few measurements, without the need of performing full-state tomography. Moreover, our method can be directly applied to other platforms using different degrees of freedom and other steering criteria, such as for generalized entropy \cite{zhu16prl-renyi,costa18pra-tsallis}. 
	
	Since there is a realistic trade-off between dimensionality and noise robustness in HD quantum information \cite{zhu19arxiv-noise,marcus19prx-noise}, it is curious to investigate the trade-off between number of measurement settings and noise robustness. Interestingly, we found when the discrepancy between detective efficiencies on different measurement bases is not large, the noise robustness could be always enhanced by increasing measurement settings.
	
	Finally, another interest is to develop multisetting methods for noise-robust detection of genuine HD steering, as quantified by the Schmidt
	number~\cite{designolle2020genuine}. In addition, noisy steerable states provide potential applications in the task of subchannel discrimination \cite{piani09prl-channel,subchannel-watrous2015prl}. 
	\\[2pt]	
	
	%\section*{Acknowledgments}	
	This work was supported by the National Nature Science Foundation of China (Grant Nos. 91736104, 12174301, 11804271, 12074307, 11534008 and 92050103),  China Postdoctoral Science Foundation via Project No. 2020M673366.
	\\[2pt]
		
	\textit{Note added.---}During the submission of the paper, we became aware of a recent related work that introduces a single-outcome HD steering inequality that is tolerant to noise and loss \cite{22arxiv-malik-noise}.

\nocite{*}
\bibliography{bib.bib}

%apsrev4-2.bst 2019-01-14 (MD) hand-edited version of apsrev4-1.bst
%Control: key (0)
%Control: author (8) initials jnrlst
%Control: editor formatted (1) identically to author
%Control: production of article title (0) allowed
%Control: page (0) single
%Control: year (1) truncated
%Control: production of eprint (0) enabled
\begin{thebibliography}{45}%
\makeatletter
\providecommand \@ifxundefined [1]{%
 \@ifx{#1\undefined}
}%
\providecommand \@ifnum [1]{%
 \ifnum #1\expandafter \@firstoftwo
 \else \expandafter \@secondoftwo
 \fi
}%
\providecommand \@ifx [1]{%
 \ifx #1\expandafter \@firstoftwo
 \else \expandafter \@secondoftwo
 \fi
}%
\providecommand \natexlab [1]{#1}%
\providecommand \enquote  [1]{``#1''}%
\providecommand \bibnamefont  [1]{#1}%
\providecommand \bibfnamefont [1]{#1}%
\providecommand \citenamefont [1]{#1}%
\providecommand \href@noop [0]{\@secondoftwo}%
\providecommand \href [0]{\begingroup \@sanitize@url \@href}%
\providecommand \@href[1]{\@@startlink{#1}\@@href}%
\providecommand \@@href[1]{\endgroup#1\@@endlink}%
\providecommand \@sanitize@url [0]{\catcode `\\12\catcode `\$12\catcode
  `\&12\catcode `\#12\catcode `\^12\catcode `\_12\catcode `\%12\relax}%
\providecommand \@@startlink[1]{}%
\providecommand \@@endlink[0]{}%
\providecommand \url  [0]{\begingroup\@sanitize@url \@url }%
\providecommand \@url [1]{\endgroup\@href {#1}{\urlprefix }}%
\providecommand \urlprefix  [0]{URL }%
\providecommand \Eprint [0]{\href }%
\providecommand \doibase [0]{https://doi.org/}%
\providecommand \selectlanguage [0]{\@gobble}%
\providecommand \bibinfo  [0]{\@secondoftwo}%
\providecommand \bibfield  [0]{\@secondoftwo}%
\providecommand \translation [1]{[#1]}%
\providecommand \BibitemOpen [0]{}%
\providecommand \bibitemStop [0]{}%
\providecommand \bibitemNoStop [0]{.\EOS\space}%
\providecommand \EOS [0]{\spacefactor3000\relax}%
\providecommand \BibitemShut  [1]{\csname bibitem#1\endcsname}%
\let\auto@bib@innerbib\@empty
%</preamble>
\bibitem [{\citenamefont {Wiseman}\ \emph {et~al.}(2007)\citenamefont
  {Wiseman}, \citenamefont {Jones},\ and\ \citenamefont
  {Doherty}}]{wiseman2007prl}%
  \BibitemOpen
  \bibfield  {author} {\bibinfo {author} {\bibfnamefont {H.~M.}\ \bibnamefont
  {Wiseman}}, \bibinfo {author} {\bibfnamefont {S.~J.}\ \bibnamefont {Jones}},\
  and\ \bibinfo {author} {\bibfnamefont {A.~C.}\ \bibnamefont {Doherty}},\
  }\bibfield  {title} {\bibinfo {title} {Steering, entanglement, nonlocality,
  and the einstein-podolsky-rosen paradox},\ }\href
  {https://doi.org/10.1103/PhysRevLett.98.140402} {\bibfield  {journal}
  {\bibinfo  {journal} {Phys. Rev. Lett.}\ }\textbf {\bibinfo {volume} {98}},\
  \bibinfo {pages} {140402} (\bibinfo {year} {2007})}\BibitemShut {NoStop}%
\bibitem [{\citenamefont {Quintino}\ \emph {et~al.}(2015)\citenamefont
  {Quintino}, \citenamefont {V\'ertesi}, \citenamefont {Cavalcanti},
  \citenamefont {Augusiak}, \citenamefont {Demianowicz}, \citenamefont
  {Ac\'{\i}n},\ and\ \citenamefont {Brunner}}]{quintino2015inequivalence}%
  \BibitemOpen
  \bibfield  {author} {\bibinfo {author} {\bibfnamefont {M.~T.}\ \bibnamefont
  {Quintino}}, \bibinfo {author} {\bibfnamefont {T.}~\bibnamefont {V\'ertesi}},
  \bibinfo {author} {\bibfnamefont {D.}~\bibnamefont {Cavalcanti}}, \bibinfo
  {author} {\bibfnamefont {R.}~\bibnamefont {Augusiak}}, \bibinfo {author}
  {\bibfnamefont {M.}~\bibnamefont {Demianowicz}}, \bibinfo {author}
  {\bibfnamefont {A.}~\bibnamefont {Ac\'{\i}n}},\ and\ \bibinfo {author}
  {\bibfnamefont {N.}~\bibnamefont {Brunner}},\ }\bibfield  {title} {\bibinfo
  {title} {Inequivalence of entanglement, steering, and bell nonlocality for
  general measurements},\ }\href {https://doi.org/10.1103/PhysRevA.92.032107}
  {\bibfield  {journal} {\bibinfo  {journal} {Phys. Rev. A}\ }\textbf {\bibinfo
  {volume} {92}},\ \bibinfo {pages} {032107} (\bibinfo {year}
  {2015})}\BibitemShut {NoStop}%
\bibitem [{\citenamefont {Einstein}\ \emph {et~al.}(1935)\citenamefont
  {Einstein}, \citenamefont {Podolsky},\ and\ \citenamefont
  {Rosen}}]{einstein1935can}%
  \BibitemOpen
  \bibfield  {author} {\bibinfo {author} {\bibfnamefont {A.}~\bibnamefont
  {Einstein}}, \bibinfo {author} {\bibfnamefont {B.}~\bibnamefont {Podolsky}},\
  and\ \bibinfo {author} {\bibfnamefont {N.}~\bibnamefont {Rosen}},\ }\bibfield
   {title} {\bibinfo {title} {Can quantum-mechanical description of physical
  reality be considered complete?},\ }\href
  {https://link.aps.org/doi/10.1103/PhysRev.47.777} {\bibfield  {journal}
  {\bibinfo  {journal} {Phys. Rev.}\ }\textbf {\bibinfo {volume} {47}},\
  \bibinfo {pages} {777} (\bibinfo {year} {1935})}\BibitemShut {NoStop}%
\bibitem [{\citenamefont {Schrödinger}(1935)}]{schrodinger1935}%
  \BibitemOpen
  \bibfield  {author} {\bibinfo {author} {\bibfnamefont {E.}~\bibnamefont
  {Schrödinger}},\ }\bibfield  {title} {\bibinfo {title} {Discussion of
  probability relations between separated systems},\ }\href
  {https://doi.org/10.1017/S0305004100013554} {\bibfield  {journal} {\bibinfo
  {journal} {Mathematical Proceedings of the Cambridge Philosophical Society}\
  }\textbf {\bibinfo {volume} {31}},\ \bibinfo {pages} {555–563} (\bibinfo
  {year} {1935})}\BibitemShut {NoStop}%
\bibitem [{\citenamefont {Uola}\ \emph {et~al.}(2020)\citenamefont {Uola},
  \citenamefont {Costa}, \citenamefont {Nguyen},\ and\ \citenamefont
  {G\"uhne}}]{uola20RMP-review}%
  \BibitemOpen
  \bibfield  {author} {\bibinfo {author} {\bibfnamefont {R.}~\bibnamefont
  {Uola}}, \bibinfo {author} {\bibfnamefont {A.~C.~S.}\ \bibnamefont {Costa}},
  \bibinfo {author} {\bibfnamefont {H.~C.}\ \bibnamefont {Nguyen}},\ and\
  \bibinfo {author} {\bibfnamefont {O.}~\bibnamefont {G\"uhne}},\ }\bibfield
  {title} {\bibinfo {title} {Quantum steering},\ }\href
  {https://doi.org/10.1103/RevModPhys.92.015001} {\bibfield  {journal}
  {\bibinfo  {journal} {Rev. Mod. Phys.}\ }\textbf {\bibinfo {volume} {92}},\
  \bibinfo {pages} {015001} (\bibinfo {year} {2020})}\BibitemShut {NoStop}%
\bibitem [{\citenamefont {Cavalcanti}\ and\ \citenamefont
  {Skrzypczyk}(2016)}]{16review-steering-cavalcanti2016quantum}%
  \BibitemOpen
  \bibfield  {author} {\bibinfo {author} {\bibfnamefont {D.}~\bibnamefont
  {Cavalcanti}}\ and\ \bibinfo {author} {\bibfnamefont {P.}~\bibnamefont
  {Skrzypczyk}},\ }\bibfield  {title} {\bibinfo {title} {Quantum steering: a
  review with focus on semidefinite programming},\ }\href
  {https://doi.org/10.1088/1361-6633/80/2/024001} {\bibfield  {journal}
  {\bibinfo  {journal} {Reports on Progress in Physics}\ }\textbf {\bibinfo
  {volume} {80}},\ \bibinfo {pages} {024001} (\bibinfo {year}
  {2016})}\BibitemShut {NoStop}%
\bibitem [{\citenamefont {Reid}\ \emph {et~al.}(2009)\citenamefont {Reid},
  \citenamefont {Drummond}, \citenamefont {Bowen}, \citenamefont {Cavalcanti},
  \citenamefont {Lam}, \citenamefont {Bachor}, \citenamefont {Andersen},\ and\
  \citenamefont {Leuchs}}]{09review-steering-reid2009colloquium}%
  \BibitemOpen
  \bibfield  {author} {\bibinfo {author} {\bibfnamefont {M.~D.}\ \bibnamefont
  {Reid}}, \bibinfo {author} {\bibfnamefont {P.~D.}\ \bibnamefont {Drummond}},
  \bibinfo {author} {\bibfnamefont {W.~P.}\ \bibnamefont {Bowen}}, \bibinfo
  {author} {\bibfnamefont {E.~G.}\ \bibnamefont {Cavalcanti}}, \bibinfo
  {author} {\bibfnamefont {P.~K.}\ \bibnamefont {Lam}}, \bibinfo {author}
  {\bibfnamefont {H.~A.}\ \bibnamefont {Bachor}}, \bibinfo {author}
  {\bibfnamefont {U.~L.}\ \bibnamefont {Andersen}},\ and\ \bibinfo {author}
  {\bibfnamefont {G.}~\bibnamefont {Leuchs}},\ }\bibfield  {title} {\bibinfo
  {title} {Colloquium: The einstein-podolsky-rosen paradox: From concepts to
  applications},\ }\href {https://doi.org/10.1103/RevModPhys.81.1727}
  {\bibfield  {journal} {\bibinfo  {journal} {Rev. Mod. Phys.}\ }\textbf
  {\bibinfo {volume} {81}},\ \bibinfo {pages} {1727} (\bibinfo {year}
  {2009})}\BibitemShut {NoStop}%
\bibitem [{\citenamefont {Piani}\ and\ \citenamefont
  {Watrous}(2015)}]{subchannel-watrous2015prl}%
  \BibitemOpen
  \bibfield  {author} {\bibinfo {author} {\bibfnamefont {M.}~\bibnamefont
  {Piani}}\ and\ \bibinfo {author} {\bibfnamefont {J.}~\bibnamefont
  {Watrous}},\ }\bibfield  {title} {\bibinfo {title} {Necessary and sufficient
  quantum information characterization of einstein-podolsky-rosen steering},\
  }\href {https://doi.org/10.1103/PhysRevLett.114.060404} {\bibfield  {journal}
  {\bibinfo  {journal} {Phys. Rev. Lett.}\ }\textbf {\bibinfo {volume} {114}},\
  \bibinfo {pages} {060404} (\bibinfo {year} {2015})}\BibitemShut {NoStop}%
\bibitem [{\citenamefont {Sun}\ \emph {et~al.}(2018)\citenamefont {Sun},
  \citenamefont {Ye}, \citenamefont {Xiao}, \citenamefont {Xu}, \citenamefont
  {Wu}, \citenamefont {Xu}, \citenamefont {Chen}, \citenamefont {Li},\ and\
  \citenamefont {Guo}}]{subchannel-sun2018npj}%
  \BibitemOpen
  \bibfield  {author} {\bibinfo {author} {\bibfnamefont {K.}~\bibnamefont
  {Sun}}, \bibinfo {author} {\bibfnamefont {X.-J.}\ \bibnamefont {Ye}},
  \bibinfo {author} {\bibfnamefont {Y.}~\bibnamefont {Xiao}}, \bibinfo {author}
  {\bibfnamefont {X.-Y.}\ \bibnamefont {Xu}}, \bibinfo {author} {\bibfnamefont
  {Y.-C.}\ \bibnamefont {Wu}}, \bibinfo {author} {\bibfnamefont {J.-S.}\
  \bibnamefont {Xu}}, \bibinfo {author} {\bibfnamefont {J.-L.}\ \bibnamefont
  {Chen}}, \bibinfo {author} {\bibfnamefont {C.-F.}\ \bibnamefont {Li}},\ and\
  \bibinfo {author} {\bibfnamefont {G.-C.}\ \bibnamefont {Guo}},\ }\bibfield
  {title} {\bibinfo {title} {Demonstration of einstein--podolsky--rosen
  steering with enhanced subchannel discrimination},\ }\href
  {https://doi.org/https://doi.org/10.1038/s41534-018-0067-1} {\bibfield
  {journal} {\bibinfo  {journal} {npj Quantum Information}\ }\textbf {\bibinfo
  {volume} {4}},\ \bibinfo {pages} {1} (\bibinfo {year} {2018})}\BibitemShut
  {NoStop}%
\bibitem [{\citenamefont {Branciard}\ \emph {et~al.}(2012)\citenamefont
  {Branciard}, \citenamefont {Cavalcanti}, \citenamefont {Walborn},
  \citenamefont {Scarani},\ and\ \citenamefont
  {Wiseman}}]{Wiseman12pra-SDI-QKD}%
  \BibitemOpen
  \bibfield  {author} {\bibinfo {author} {\bibfnamefont {C.}~\bibnamefont
  {Branciard}}, \bibinfo {author} {\bibfnamefont {E.~G.}\ \bibnamefont
  {Cavalcanti}}, \bibinfo {author} {\bibfnamefont {S.~P.}\ \bibnamefont
  {Walborn}}, \bibinfo {author} {\bibfnamefont {V.}~\bibnamefont {Scarani}},\
  and\ \bibinfo {author} {\bibfnamefont {H.~M.}\ \bibnamefont {Wiseman}},\
  }\bibfield  {title} {\bibinfo {title} {One-sided device-independent quantum
  key distribution: Security, feasibility, and the connection with steering},\
  }\href {https://doi.org/10.1103/PhysRevA.85.010301} {\bibfield  {journal}
  {\bibinfo  {journal} {Phys. Rev. A}\ }\textbf {\bibinfo {volume} {85}},\
  \bibinfo {pages} {010301} (\bibinfo {year} {2012})}\BibitemShut {NoStop}%
\bibitem [{\citenamefont {He}\ \emph {et~al.}(2015)\citenamefont {He},
  \citenamefont {Rosales-Z\'arate}, \citenamefont {Adesso},\ and\ \citenamefont
  {Reid}}]{tele-he2015prl}%
  \BibitemOpen
  \bibfield  {author} {\bibinfo {author} {\bibfnamefont {Q.}~\bibnamefont
  {He}}, \bibinfo {author} {\bibfnamefont {L.}~\bibnamefont
  {Rosales-Z\'arate}}, \bibinfo {author} {\bibfnamefont {G.}~\bibnamefont
  {Adesso}},\ and\ \bibinfo {author} {\bibfnamefont {M.~D.}\ \bibnamefont
  {Reid}},\ }\bibfield  {title} {\bibinfo {title} {Secure continuous variable
  teleportation and einstein-podolsky-rosen steering},\ }\href
  {https://doi.org/10.1103/PhysRevLett.115.180502} {\bibfield  {journal}
  {\bibinfo  {journal} {Phys. Rev. Lett.}\ }\textbf {\bibinfo {volume} {115}},\
  \bibinfo {pages} {180502} (\bibinfo {year} {2015})}\BibitemShut {NoStop}%
\bibitem [{\citenamefont {Reid}(2013)}]{tele-reid2013pra}%
  \BibitemOpen
  \bibfield  {author} {\bibinfo {author} {\bibfnamefont {M.~D.}\ \bibnamefont
  {Reid}},\ }\bibfield  {title} {\bibinfo {title} {Signifying quantum
  benchmarks for qubit teleportation and secure quantum communication using
  einstein-podolsky-rosen steering inequalities},\ }\href
  {https://doi.org/10.1103/PhysRevA.88.062338} {\bibfield  {journal} {\bibinfo
  {journal} {Phys. Rev. A}\ }\textbf {\bibinfo {volume} {88}},\ \bibinfo
  {pages} {062338} (\bibinfo {year} {2013})}\BibitemShut {NoStop}%
\bibitem [{\citenamefont {Law}\ \emph {et~al.}(2014)\citenamefont {Law},
  \citenamefont {Thinh}, \citenamefont {Bancal},\ and\ \citenamefont
  {Scarani}}]{random-law2014quantum}%
  \BibitemOpen
  \bibfield  {author} {\bibinfo {author} {\bibfnamefont {Y.~Z.}\ \bibnamefont
  {Law}}, \bibinfo {author} {\bibfnamefont {L.~P.}\ \bibnamefont {Thinh}},
  \bibinfo {author} {\bibfnamefont {J.-D.}\ \bibnamefont {Bancal}},\ and\
  \bibinfo {author} {\bibfnamefont {V.}~\bibnamefont {Scarani}},\ }\bibfield
  {title} {\bibinfo {title} {Quantum randomness extraction for various levels
  of characterization of the devices},\ }\href
  {https://doi.org/10.1088/1751-8113/47/42/424028} {\bibfield  {journal}
  {\bibinfo  {journal} {Journal of Physics A: Mathematical and Theoretical}\
  }\textbf {\bibinfo {volume} {47}},\ \bibinfo {pages} {424028} (\bibinfo
  {year} {2014})}\BibitemShut {NoStop}%
\bibitem [{\citenamefont {Skrzypczyk}\ and\ \citenamefont
  {Cavalcanti}(2018)}]{skrzypczyk2018prlmaximal}%
  \BibitemOpen
  \bibfield  {author} {\bibinfo {author} {\bibfnamefont {P.}~\bibnamefont
  {Skrzypczyk}}\ and\ \bibinfo {author} {\bibfnamefont {D.}~\bibnamefont
  {Cavalcanti}},\ }\bibfield  {title} {\bibinfo {title} {Maximal randomness
  generation from steering inequality violations using qudits},\ }\href
  {https://doi.org/10.1103/PhysRevLett.120.260401} {\bibfield  {journal}
  {\bibinfo  {journal} {Phys. Rev. Lett.}\ }\textbf {\bibinfo {volume} {120}},\
  \bibinfo {pages} {260401} (\bibinfo {year} {2018})}\BibitemShut {NoStop}%
\bibitem [{\citenamefont {Guo}\ \emph {et~al.}(2019)\citenamefont {Guo},
  \citenamefont {Cheng}, \citenamefont {Hu}, \citenamefont {Liu}, \citenamefont
  {Huang}, \citenamefont {Huang}, \citenamefont {Li}, \citenamefont {Guo},\
  and\ \citenamefont {Cavalcanti}}]{guo2019prl-random}%
  \BibitemOpen
  \bibfield  {author} {\bibinfo {author} {\bibfnamefont {Y.}~\bibnamefont
  {Guo}}, \bibinfo {author} {\bibfnamefont {S.}~\bibnamefont {Cheng}}, \bibinfo
  {author} {\bibfnamefont {X.}~\bibnamefont {Hu}}, \bibinfo {author}
  {\bibfnamefont {B.-H.}\ \bibnamefont {Liu}}, \bibinfo {author} {\bibfnamefont
  {E.-M.}\ \bibnamefont {Huang}}, \bibinfo {author} {\bibfnamefont {Y.-F.}\
  \bibnamefont {Huang}}, \bibinfo {author} {\bibfnamefont {C.-F.}\ \bibnamefont
  {Li}}, \bibinfo {author} {\bibfnamefont {G.-C.}\ \bibnamefont {Guo}},\ and\
  \bibinfo {author} {\bibfnamefont {E.~G.}\ \bibnamefont {Cavalcanti}},\
  }\bibfield  {title} {\bibinfo {title} {Experimental
  measurement-device-independent quantum steering and randomness generation
  beyond qubits},\ }\href {https://doi.org/10.1103/PhysRevLett.123.170402}
  {\bibfield  {journal} {\bibinfo  {journal} {Phys. Rev. Lett.}\ }\textbf
  {\bibinfo {volume} {123}},\ \bibinfo {pages} {170402} (\bibinfo {year}
  {2019})}\BibitemShut {NoStop}%
\bibitem [{\citenamefont {Wang}\ \emph {et~al.}(2018)\citenamefont {Wang},
  \citenamefont {Paesani}, \citenamefont {Ding}, \citenamefont {Santagati},
  \citenamefont {Skrzypczyk}, \citenamefont {Salavrakos}, \citenamefont {Tura},
  \citenamefont {Augusiak}, \citenamefont {Man{\v{c}}inska}, \citenamefont
  {Bacco} \emph {et~al.}}]{wang2018science}%
  \BibitemOpen
  \bibfield  {author} {\bibinfo {author} {\bibfnamefont {J.}~\bibnamefont
  {Wang}}, \bibinfo {author} {\bibfnamefont {S.}~\bibnamefont {Paesani}},
  \bibinfo {author} {\bibfnamefont {Y.}~\bibnamefont {Ding}}, \bibinfo {author}
  {\bibfnamefont {R.}~\bibnamefont {Santagati}}, \bibinfo {author}
  {\bibfnamefont {P.}~\bibnamefont {Skrzypczyk}}, \bibinfo {author}
  {\bibfnamefont {A.}~\bibnamefont {Salavrakos}}, \bibinfo {author}
  {\bibfnamefont {J.}~\bibnamefont {Tura}}, \bibinfo {author} {\bibfnamefont
  {R.}~\bibnamefont {Augusiak}}, \bibinfo {author} {\bibfnamefont
  {L.}~\bibnamefont {Man{\v{c}}inska}}, \bibinfo {author} {\bibfnamefont
  {D.}~\bibnamefont {Bacco}}, \emph {et~al.},\ }\bibfield  {title} {\bibinfo
  {title} {Multidimensional quantum entanglement with large-scale integrated
  optics},\ }\href {https://doi.org/10.1126/science.aar7053} {\bibfield
  {journal} {\bibinfo  {journal} {Science}\ }\textbf {\bibinfo {volume}
  {360}},\ \bibinfo {pages} {285} (\bibinfo {year} {2018})}\BibitemShut
  {NoStop}%
\bibitem [{\citenamefont {Cozzolino}\ \emph {et~al.}(2019)\citenamefont
  {Cozzolino}, \citenamefont {Da~Lio}, \citenamefont {Bacco},\ and\
  \citenamefont {Oxenl{\o}we}}]{cozzolino2019-HDreview}%
  \BibitemOpen
  \bibfield  {author} {\bibinfo {author} {\bibfnamefont {D.}~\bibnamefont
  {Cozzolino}}, \bibinfo {author} {\bibfnamefont {B.}~\bibnamefont {Da~Lio}},
  \bibinfo {author} {\bibfnamefont {D.}~\bibnamefont {Bacco}},\ and\ \bibinfo
  {author} {\bibfnamefont {L.~K.}\ \bibnamefont {Oxenl{\o}we}},\ }\bibfield
  {title} {\bibinfo {title} {High-dimensional quantum communication: Benefits,
  progress, and future challenges},\ }\href
  {https://doi.org/https://doi.org/10.1002/qute.201900038} {\bibfield
  {journal} {\bibinfo  {journal} {Advanced Quantum Technologies}\ }\textbf
  {\bibinfo {volume} {2}},\ \bibinfo {pages} {1900038} (\bibinfo {year}
  {2019})}\BibitemShut {NoStop}%
\bibitem [{\citenamefont {Vaziri}\ \emph {et~al.}(2002)\citenamefont {Vaziri},
  \citenamefont {Weihs},\ and\ \citenamefont {Zeilinger}}]{zeilinger02prl}%
  \BibitemOpen
  \bibfield  {author} {\bibinfo {author} {\bibfnamefont {A.}~\bibnamefont
  {Vaziri}}, \bibinfo {author} {\bibfnamefont {G.}~\bibnamefont {Weihs}},\ and\
  \bibinfo {author} {\bibfnamefont {A.}~\bibnamefont {Zeilinger}},\ }\bibfield
  {title} {\bibinfo {title} {Experimental two-photon, three-dimensional
  entanglement for quantum communication},\ }\href
  {https://doi.org/10.1103/PhysRevLett.89.240401} {\bibfield  {journal}
  {\bibinfo  {journal} {Phys. Rev. Lett.}\ }\textbf {\bibinfo {volume} {89}},\
  \bibinfo {pages} {240401} (\bibinfo {year} {2002})}\BibitemShut {NoStop}%
\bibitem [{\citenamefont {Thew}\ \emph {et~al.}(2004)\citenamefont {Thew},
  \citenamefont {Ac\'{\i}n}, \citenamefont {Zbinden},\ and\ \citenamefont
  {Gisin}}]{thew2004prl-bell}%
  \BibitemOpen
  \bibfield  {author} {\bibinfo {author} {\bibfnamefont {R.~T.}\ \bibnamefont
  {Thew}}, \bibinfo {author} {\bibfnamefont {A.}~\bibnamefont {Ac\'{\i}n}},
  \bibinfo {author} {\bibfnamefont {H.}~\bibnamefont {Zbinden}},\ and\ \bibinfo
  {author} {\bibfnamefont {N.}~\bibnamefont {Gisin}},\ }\bibfield  {title}
  {\bibinfo {title} {Bell-type test of energy-time entangled qutrits},\ }\href
  {https://doi.org/10.1103/PhysRevLett.93.010503} {\bibfield  {journal}
  {\bibinfo  {journal} {Phys. Rev. Lett.}\ }\textbf {\bibinfo {volume} {93}},\
  \bibinfo {pages} {010503} (\bibinfo {year} {2004})}\BibitemShut {NoStop}%
\bibitem [{\citenamefont {Gröblacher}\ \emph {et~al.}(2006)\citenamefont
  {Gröblacher}, \citenamefont {Jennewein}, \citenamefont {Vaziri},
  \citenamefont {Weihs},\ and\ \citenamefont
  {Zeilinger}}]{zeilinger2006-capacity}%
  \BibitemOpen
  \bibfield  {author} {\bibinfo {author} {\bibfnamefont {S.}~\bibnamefont
  {Gröblacher}}, \bibinfo {author} {\bibfnamefont {T.}~\bibnamefont
  {Jennewein}}, \bibinfo {author} {\bibfnamefont {A.}~\bibnamefont {Vaziri}},
  \bibinfo {author} {\bibfnamefont {G.}~\bibnamefont {Weihs}},\ and\ \bibinfo
  {author} {\bibfnamefont {A.}~\bibnamefont {Zeilinger}},\ }\bibfield  {title}
  {\bibinfo {title} {Experimental quantum cryptography with qutrits},\ }\href
  {https://doi.org/10.1088/1367-2630/8/5/075} {\bibfield  {journal} {\bibinfo
  {journal} {New Journal of Physics}\ }\textbf {\bibinfo {volume} {8}},\
  \bibinfo {pages} {75} (\bibinfo {year} {2006})}\BibitemShut {NoStop}%
\bibitem [{\citenamefont {Hu}\ \emph {et~al.}(2018)\citenamefont {Hu},
  \citenamefont {Guo}, \citenamefont {Liu}, \citenamefont {Huang},
  \citenamefont {Li},\ and\ \citenamefont {Guo}}]{hu2018beating}%
  \BibitemOpen
  \bibfield  {author} {\bibinfo {author} {\bibfnamefont {X.-M.}\ \bibnamefont
  {Hu}}, \bibinfo {author} {\bibfnamefont {Y.}~\bibnamefont {Guo}}, \bibinfo
  {author} {\bibfnamefont {B.-H.}\ \bibnamefont {Liu}}, \bibinfo {author}
  {\bibfnamefont {Y.-F.}\ \bibnamefont {Huang}}, \bibinfo {author}
  {\bibfnamefont {C.-F.}\ \bibnamefont {Li}},\ and\ \bibinfo {author}
  {\bibfnamefont {G.-C.}\ \bibnamefont {Guo}},\ }\bibfield  {title} {\bibinfo
  {title} {Beating the channel capacity limit for superdense coding with
  entangled ququarts},\ }\href {https://doi.org/10.1126/sciadv.aat9304}
  {\bibfield  {journal} {\bibinfo  {journal} {Science Advances}\ }\textbf
  {\bibinfo {volume} {4}},\ \bibinfo {pages} {eaat9304} (\bibinfo {year}
  {2018})}\BibitemShut {NoStop}%
\bibitem [{\citenamefont {Islam}\ \emph {et~al.}(2017)\citenamefont {Islam},
  \citenamefont {Lim}, \citenamefont {Cahall}, \citenamefont {Kim},\ and\
  \citenamefont {Gauthier}}]{islam2017sa-provably}%
  \BibitemOpen
  \bibfield  {author} {\bibinfo {author} {\bibfnamefont {N.~T.}\ \bibnamefont
  {Islam}}, \bibinfo {author} {\bibfnamefont {C.~C.~W.}\ \bibnamefont {Lim}},
  \bibinfo {author} {\bibfnamefont {C.}~\bibnamefont {Cahall}}, \bibinfo
  {author} {\bibfnamefont {J.}~\bibnamefont {Kim}},\ and\ \bibinfo {author}
  {\bibfnamefont {D.~J.}\ \bibnamefont {Gauthier}},\ }\bibfield  {title}
  {\bibinfo {title} {Provably secure and high-rate quantum key distribution
  with time-bin qudits},\ }\href {https://doi.org/10.1126/sciadv.1701491}
  {\bibfield  {journal} {\bibinfo  {journal} {Science Advances}\ }\textbf
  {\bibinfo {volume} {3}},\ \bibinfo {pages} {e1701491} (\bibinfo {year}
  {2017})}\BibitemShut {NoStop}%
\bibitem [{\citenamefont {Ecker}\ \emph {et~al.}(2019)\citenamefont {Ecker},
  \citenamefont {Bouchard}, \citenamefont {Bulla}, \citenamefont {Brandt},
  \citenamefont {Kohout}, \citenamefont {Steinlechner}, \citenamefont
  {Fickler}, \citenamefont {Malik}, \citenamefont {Guryanova}, \citenamefont
  {Ursin},\ and\ \citenamefont {Huber}}]{marcus19prx-noise}%
  \BibitemOpen
  \bibfield  {author} {\bibinfo {author} {\bibfnamefont {S.}~\bibnamefont
  {Ecker}}, \bibinfo {author} {\bibfnamefont {F.}~\bibnamefont {Bouchard}},
  \bibinfo {author} {\bibfnamefont {L.}~\bibnamefont {Bulla}}, \bibinfo
  {author} {\bibfnamefont {F.}~\bibnamefont {Brandt}}, \bibinfo {author}
  {\bibfnamefont {O.}~\bibnamefont {Kohout}}, \bibinfo {author} {\bibfnamefont
  {F.}~\bibnamefont {Steinlechner}}, \bibinfo {author} {\bibfnamefont
  {R.}~\bibnamefont {Fickler}}, \bibinfo {author} {\bibfnamefont
  {M.}~\bibnamefont {Malik}}, \bibinfo {author} {\bibfnamefont
  {Y.}~\bibnamefont {Guryanova}}, \bibinfo {author} {\bibfnamefont
  {R.}~\bibnamefont {Ursin}},\ and\ \bibinfo {author} {\bibfnamefont
  {M.}~\bibnamefont {Huber}},\ }\bibfield  {title} {\bibinfo {title}
  {Overcoming noise in entanglement distribution},\ }\href
  {https://doi.org/10.1103/PhysRevX.9.041042} {\bibfield  {journal} {\bibinfo
  {journal} {Phys. Rev. X}\ }\textbf {\bibinfo {volume} {9}},\ \bibinfo {pages}
  {041042} (\bibinfo {year} {2019})}\BibitemShut {NoStop}%
\bibitem [{\citenamefont {V\'ertesi}\ \emph {et~al.}(2010)\citenamefont
  {V\'ertesi}, \citenamefont {Pironio},\ and\ \citenamefont
  {Brunner}}]{vertesi2010prl-noise}%
  \BibitemOpen
  \bibfield  {author} {\bibinfo {author} {\bibfnamefont {T.}~\bibnamefont
  {V\'ertesi}}, \bibinfo {author} {\bibfnamefont {S.}~\bibnamefont {Pironio}},\
  and\ \bibinfo {author} {\bibfnamefont {N.}~\bibnamefont {Brunner}},\
  }\bibfield  {title} {\bibinfo {title} {Closing the detection loophole in bell
  experiments using qudits},\ }\href
  {https://doi.org/10.1103/PhysRevLett.104.060401} {\bibfield  {journal}
  {\bibinfo  {journal} {Phys. Rev. Lett.}\ }\textbf {\bibinfo {volume} {104}},\
  \bibinfo {pages} {060401} (\bibinfo {year} {2010})}\BibitemShut {NoStop}%
\bibitem [{\citenamefont {Zhu}\ \emph {et~al.}(2021)\citenamefont {Zhu},
  \citenamefont {Tyler}, \citenamefont {Valencia}, \citenamefont {Malik},\ and\
  \citenamefont {Leach}}]{zhu19arxiv-noise}%
  \BibitemOpen
  \bibfield  {author} {\bibinfo {author} {\bibfnamefont {F.}~\bibnamefont
  {Zhu}}, \bibinfo {author} {\bibfnamefont {M.}~\bibnamefont {Tyler}}, \bibinfo
  {author} {\bibfnamefont {N.~H.}\ \bibnamefont {Valencia}}, \bibinfo {author}
  {\bibfnamefont {M.}~\bibnamefont {Malik}},\ and\ \bibinfo {author}
  {\bibfnamefont {J.}~\bibnamefont {Leach}},\ }\bibfield  {title} {\bibinfo
  {title} {Is high-dimensional photonic entanglement robust to noise?},\ }\href
  {https://doi.org/https://doi.org/10.1116/5.0033889} {\bibfield  {journal}
  {\bibinfo  {journal} {AVS Quantum Science}\ }\textbf {\bibinfo {volume}
  {3}},\ \bibinfo {pages} {011401} (\bibinfo {year} {2021})}\BibitemShut
  {NoStop}%
\bibitem [{\citenamefont {Designolle}\ \emph {et~al.}(2021)\citenamefont
  {Designolle}, \citenamefont {Srivastav}, \citenamefont {Uola}, \citenamefont
  {Valencia}, \citenamefont {McCutcheon}, \citenamefont {Malik},\ and\
  \citenamefont {Brunner}}]{designolle2020genuine}%
  \BibitemOpen
  \bibfield  {author} {\bibinfo {author} {\bibfnamefont {S.}~\bibnamefont
  {Designolle}}, \bibinfo {author} {\bibfnamefont {V.}~\bibnamefont
  {Srivastav}}, \bibinfo {author} {\bibfnamefont {R.}~\bibnamefont {Uola}},
  \bibinfo {author} {\bibfnamefont {N.~H.}\ \bibnamefont {Valencia}}, \bibinfo
  {author} {\bibfnamefont {W.}~\bibnamefont {McCutcheon}}, \bibinfo {author}
  {\bibfnamefont {M.}~\bibnamefont {Malik}},\ and\ \bibinfo {author}
  {\bibfnamefont {N.}~\bibnamefont {Brunner}},\ }\bibfield  {title} {\bibinfo
  {title} {Genuine high-dimensional quantum steering},\ }\href
  {https://doi.org/10.1103/PhysRevLett.126.200404} {\bibfield  {journal}
  {\bibinfo  {journal} {Phys. Rev. Lett.}\ }\textbf {\bibinfo {volume} {126}},\
  \bibinfo {pages} {200404} (\bibinfo {year} {2021})}\BibitemShut {NoStop}%
\bibitem [{\citenamefont {Zeng}\ \emph {et~al.}(2018)\citenamefont {Zeng},
  \citenamefont {Wang}, \citenamefont {Li},\ and\ \citenamefont
  {Zhang}}]{zeng2018prl-HDsteering}%
  \BibitemOpen
  \bibfield  {author} {\bibinfo {author} {\bibfnamefont {Q.}~\bibnamefont
  {Zeng}}, \bibinfo {author} {\bibfnamefont {B.}~\bibnamefont {Wang}}, \bibinfo
  {author} {\bibfnamefont {P.}~\bibnamefont {Li}},\ and\ \bibinfo {author}
  {\bibfnamefont {X.}~\bibnamefont {Zhang}},\ }\bibfield  {title} {\bibinfo
  {title} {Experimental high-dimensional einstein-podolsky-rosen steering},\
  }\href {https://doi.org/10.1103/PhysRevLett.120.030401} {\bibfield  {journal}
  {\bibinfo  {journal} {Phys. Rev. Lett.}\ }\textbf {\bibinfo {volume} {120}},\
  \bibinfo {pages} {030401} (\bibinfo {year} {2018})}\BibitemShut {NoStop}%
\bibitem [{\citenamefont {Li}\ \emph {et~al.}(2015)\citenamefont {Li},
  \citenamefont {Chen}, \citenamefont {Chen}, \citenamefont {Zhang},
  \citenamefont {Chen},\ and\ \citenamefont {Pan}}]{li2015PRLgenuine}%
  \BibitemOpen
  \bibfield  {author} {\bibinfo {author} {\bibfnamefont {C.-M.}\ \bibnamefont
  {Li}}, \bibinfo {author} {\bibfnamefont {K.}~\bibnamefont {Chen}}, \bibinfo
  {author} {\bibfnamefont {Y.-N.}\ \bibnamefont {Chen}}, \bibinfo {author}
  {\bibfnamefont {Q.}~\bibnamefont {Zhang}}, \bibinfo {author} {\bibfnamefont
  {Y.-A.}\ \bibnamefont {Chen}},\ and\ \bibinfo {author} {\bibfnamefont
  {J.-W.}\ \bibnamefont {Pan}},\ }\bibfield  {title} {\bibinfo {title} {Genuine
  high-order einstein-podolsky-rosen steering},\ }\href
  {https://doi.org/10.1103/PhysRevLett.115.010402} {\bibfield  {journal}
  {\bibinfo  {journal} {Phys. Rev. Lett.}\ }\textbf {\bibinfo {volume} {115}},\
  \bibinfo {pages} {010402} (\bibinfo {year} {2015})}\BibitemShut {NoStop}%
\bibitem [{\citenamefont {Marciniak}\ \emph {et~al.}(2015)\citenamefont
  {Marciniak}, \citenamefont {Rutkowski}, \citenamefont {Yin}, \citenamefont
  {Horodecki},\ and\ \citenamefont {Horodecki}}]{marciniak2015prl-unbounded}%
  \BibitemOpen
  \bibfield  {author} {\bibinfo {author} {\bibfnamefont {M.}~\bibnamefont
  {Marciniak}}, \bibinfo {author} {\bibfnamefont {A.}~\bibnamefont
  {Rutkowski}}, \bibinfo {author} {\bibfnamefont {Z.}~\bibnamefont {Yin}},
  \bibinfo {author} {\bibfnamefont {M.}~\bibnamefont {Horodecki}},\ and\
  \bibinfo {author} {\bibfnamefont {R.}~\bibnamefont {Horodecki}},\ }\bibfield
  {title} {\bibinfo {title} {Unbounded violation of quantum steering
  inequalities},\ }\href {https://doi.org/10.1103/PhysRevLett.115.170401}
  {\bibfield  {journal} {\bibinfo  {journal} {Phys. Rev. Lett.}\ }\textbf
  {\bibinfo {volume} {115}},\ \bibinfo {pages} {170401} (\bibinfo {year}
  {2015})}\BibitemShut {NoStop}%
\bibitem [{\citenamefont {Zhu}\ \emph {et~al.}(2016)\citenamefont {Zhu},
  \citenamefont {Hayashi},\ and\ \citenamefont {Chen}}]{zhu16prl-renyi}%
  \BibitemOpen
  \bibfield  {author} {\bibinfo {author} {\bibfnamefont {H.}~\bibnamefont
  {Zhu}}, \bibinfo {author} {\bibfnamefont {M.}~\bibnamefont {Hayashi}},\ and\
  \bibinfo {author} {\bibfnamefont {L.}~\bibnamefont {Chen}},\ }\bibfield
  {title} {\bibinfo {title} {Universal steering criteria},\ }\href
  {https://doi.org/10.1103/PhysRevLett.116.070403} {\bibfield  {journal}
  {\bibinfo  {journal} {Phys. Rev. Lett.}\ }\textbf {\bibinfo {volume} {116}},\
  \bibinfo {pages} {070403} (\bibinfo {year} {2016})}\BibitemShut {NoStop}%
\bibitem [{\citenamefont {Junge}\ \emph {et~al.}(2010)\citenamefont {Junge},
  \citenamefont {Palazuelos}, \citenamefont {P{\'e}rez-Garc{\'\i}a},
  \citenamefont {Villanueva},\ and\ \citenamefont {Wolf}}]{junge2010unbounded}%
  \BibitemOpen
  \bibfield  {author} {\bibinfo {author} {\bibfnamefont {M.}~\bibnamefont
  {Junge}}, \bibinfo {author} {\bibfnamefont {C.}~\bibnamefont {Palazuelos}},
  \bibinfo {author} {\bibfnamefont {D.}~\bibnamefont {P{\'e}rez-Garc{\'\i}a}},
  \bibinfo {author} {\bibfnamefont {I.}~\bibnamefont {Villanueva}},\ and\
  \bibinfo {author} {\bibfnamefont {M.~M.}\ \bibnamefont {Wolf}},\ }\bibfield
  {title} {\bibinfo {title} {Unbounded violations of bipartite bell
  inequalities via operator space theory},\ }\href
  {https://doi.org/10.1007/s00220-010-1125-5} {\bibfield  {journal} {\bibinfo
  {journal} {Communications in Mathematical Physics}\ }\textbf {\bibinfo
  {volume} {300}},\ \bibinfo {pages} {715} (\bibinfo {year}
  {2010})}\BibitemShut {NoStop}%
\bibitem [{\citenamefont {Junge}\ and\ \citenamefont
  {Palazuelos}(2011)}]{junge2011large}%
  \BibitemOpen
  \bibfield  {author} {\bibinfo {author} {\bibfnamefont {M.}~\bibnamefont
  {Junge}}\ and\ \bibinfo {author} {\bibfnamefont {C.}~\bibnamefont
  {Palazuelos}},\ }\bibfield  {title} {\bibinfo {title} {Large violation of
  bell inequalities with low entanglement},\ }\href
  {https://doi.org/10.1007/s00220-011-1296-8} {\bibfield  {journal} {\bibinfo
  {journal} {Communications in Mathematical Physics}\ }\textbf {\bibinfo
  {volume} {306}},\ \bibinfo {pages} {695} (\bibinfo {year}
  {2011})}\BibitemShut {NoStop}%
\bibitem [{\citenamefont {Ac\'{\i}n}\ \emph {et~al.}(2006)\citenamefont
  {Ac\'{\i}n}, \citenamefont {Gisin},\ and\ \citenamefont
  {Toner}}]{bell-06PRA}%
  \BibitemOpen
  \bibfield  {author} {\bibinfo {author} {\bibfnamefont {A.}~\bibnamefont
  {Ac\'{\i}n}}, \bibinfo {author} {\bibfnamefont {N.}~\bibnamefont {Gisin}},\
  and\ \bibinfo {author} {\bibfnamefont {B.}~\bibnamefont {Toner}},\ }\bibfield
   {title} {\bibinfo {title} {Grothendieck's constant and local models for
  noisy entangled quantum states},\ }\href
  {https://doi.org/10.1103/PhysRevA.73.062105} {\bibfield  {journal} {\bibinfo
  {journal} {Phys. Rev. A}\ }\textbf {\bibinfo {volume} {73}},\ \bibinfo
  {pages} {062105} (\bibinfo {year} {2006})}\BibitemShut {NoStop}%
\bibitem [{\citenamefont {Palazuelos}\ and\ \citenamefont
  {Yin}(2015)}]{yin15PRA-bellunbounded}%
  \BibitemOpen
  \bibfield  {author} {\bibinfo {author} {\bibfnamefont {C.}~\bibnamefont
  {Palazuelos}}\ and\ \bibinfo {author} {\bibfnamefont {Z.}~\bibnamefont
  {Yin}},\ }\bibfield  {title} {\bibinfo {title} {Large bipartite bell
  violations with dichotomic measurements},\ }\href
  {https://doi.org/10.1103/PhysRevA.92.052313} {\bibfield  {journal} {\bibinfo
  {journal} {Phys. Rev. A}\ }\textbf {\bibinfo {volume} {92}},\ \bibinfo
  {pages} {052313} (\bibinfo {year} {2015})}\BibitemShut {NoStop}%
\bibitem [{\citenamefont {Oppenheim}\ and\ \citenamefont
  {Wehner}(2010)}]{wehner2010uncertainty}%
  \BibitemOpen
  \bibfield  {author} {\bibinfo {author} {\bibfnamefont {J.}~\bibnamefont
  {Oppenheim}}\ and\ \bibinfo {author} {\bibfnamefont {S.}~\bibnamefont
  {Wehner}},\ }\bibfield  {title} {\bibinfo {title} {The uncertainty principle
  determines the nonlocality of quantum mechanics},\ }\href
  {https://doi.org/10.1126/science.1192065} {\bibfield  {journal} {\bibinfo
  {journal} {Science}\ }\textbf {\bibinfo {volume} {330}},\ \bibinfo {pages}
  {1072} (\bibinfo {year} {2010})}\BibitemShut {NoStop}%
\bibitem [{\citenamefont {Skrzypczyk}\ and\ \citenamefont
  {Cavalcanti}(2015)}]{skrzypczyk2015pra-loss}%
  \BibitemOpen
  \bibfield  {author} {\bibinfo {author} {\bibfnamefont {P.}~\bibnamefont
  {Skrzypczyk}}\ and\ \bibinfo {author} {\bibfnamefont {D.}~\bibnamefont
  {Cavalcanti}},\ }\bibfield  {title} {\bibinfo {title} {Loss-tolerant
  einstein-podolsky-rosen steering for arbitrary-dimensional states: Joint
  measurability and unbounded violations under losses},\ }\href
  {https://doi.org/10.1103/PhysRevA.92.022354} {\bibfield  {journal} {\bibinfo
  {journal} {Phys. Rev. A}\ }\textbf {\bibinfo {volume} {92}},\ \bibinfo
  {pages} {022354} (\bibinfo {year} {2015})}\BibitemShut {NoStop}%
\bibitem [{\citenamefont {Bengtsson}(2007)}]{bengtsson2007-mub}%
  \BibitemOpen
  \bibfield  {author} {\bibinfo {author} {\bibfnamefont {I.}~\bibnamefont
  {Bengtsson}},\ }\bibfield  {title} {\bibinfo {title} {Three ways to look at
  mutually unbiased bases},\ }\href
  {https://doi.org/https://doi.org/10.1063/1.2713445} {\bibfield  {journal}
  {\bibinfo  {journal} {AIP Conference Proceedings}\ }\textbf {\bibinfo
  {volume} {889}},\ \bibinfo {pages} {40} (\bibinfo {year} {2007})}\BibitemShut
  {NoStop}%
\bibitem [{\citenamefont {Wootters}\ and\ \citenamefont
  {Fields}(1989)}]{wootters1989-mub}%
  \BibitemOpen
  \bibfield  {author} {\bibinfo {author} {\bibfnamefont {W.~K.}\ \bibnamefont
  {Wootters}}\ and\ \bibinfo {author} {\bibfnamefont {B.~D.}\ \bibnamefont
  {Fields}},\ }\bibfield  {title} {\bibinfo {title} {Optimal
  state-determination by mutually unbiased measurements},\ }\href
  {https://doi.org/https://doi.org/10.1016/0003-4916(89)90322-9} {\bibfield
  {journal} {\bibinfo  {journal} {Annals of Physics}\ }\textbf {\bibinfo
  {volume} {191}},\ \bibinfo {pages} {363} (\bibinfo {year}
  {1989})}\BibitemShut {NoStop}%
\bibitem [{\citenamefont {Bouchard}\ \emph {et~al.}(2018)\citenamefont
  {Bouchard}, \citenamefont {Valencia}, \citenamefont {Brandt}, \citenamefont
  {Fickler}, \citenamefont {Huber},\ and\ \citenamefont {Malik}}]{18OE-ift}%
  \BibitemOpen
  \bibfield  {author} {\bibinfo {author} {\bibfnamefont {F.}~\bibnamefont
  {Bouchard}}, \bibinfo {author} {\bibfnamefont {N.~H.}\ \bibnamefont
  {Valencia}}, \bibinfo {author} {\bibfnamefont {F.}~\bibnamefont {Brandt}},
  \bibinfo {author} {\bibfnamefont {R.}~\bibnamefont {Fickler}}, \bibinfo
  {author} {\bibfnamefont {M.}~\bibnamefont {Huber}},\ and\ \bibinfo {author}
  {\bibfnamefont {M.}~\bibnamefont {Malik}},\ }\bibfield  {title} {\bibinfo
  {title} {Measuring azimuthal and radial modes of photons},\ }\href
  {https://doi.org/10.1364/OE.26.031925} {\bibfield  {journal} {\bibinfo
  {journal} {Opt. Express}\ }\textbf {\bibinfo {volume} {26}},\ \bibinfo
  {pages} {31925} (\bibinfo {year} {2018})}\BibitemShut {NoStop}%
\bibitem [{\citenamefont {Torres}\ \emph {et~al.}(2003)\citenamefont {Torres},
  \citenamefont {Alexandrescu},\ and\ \citenamefont
  {Torner}}]{torres03PRAbandwidth}%
  \BibitemOpen
  \bibfield  {author} {\bibinfo {author} {\bibfnamefont {J.~P.}\ \bibnamefont
  {Torres}}, \bibinfo {author} {\bibfnamefont {A.}~\bibnamefont
  {Alexandrescu}},\ and\ \bibinfo {author} {\bibfnamefont {L.}~\bibnamefont
  {Torner}},\ }\bibfield  {title} {\bibinfo {title} {Quantum spiral bandwidth
  of entangled two-photon states},\ }\href
  {https://doi.org/10.1103/PhysRevA.68.050301} {\bibfield  {journal} {\bibinfo
  {journal} {Phys. Rev. A}\ }\textbf {\bibinfo {volume} {68}},\ \bibinfo
  {pages} {050301} (\bibinfo {year} {2003})}\BibitemShut {NoStop}%
\bibitem [{\citenamefont {Bennett}\ \emph {et~al.}(1996)\citenamefont
  {Bennett}, \citenamefont {Bernstein}, \citenamefont {Popescu},\ and\
  \citenamefont {Schumacher}}]{bennett1996concentrating}%
  \BibitemOpen
  \bibfield  {author} {\bibinfo {author} {\bibfnamefont {C.~H.}\ \bibnamefont
  {Bennett}}, \bibinfo {author} {\bibfnamefont {H.~J.}\ \bibnamefont
  {Bernstein}}, \bibinfo {author} {\bibfnamefont {S.}~\bibnamefont {Popescu}},\
  and\ \bibinfo {author} {\bibfnamefont {B.}~\bibnamefont {Schumacher}},\
  }\bibfield  {title} {\bibinfo {title} {Concentrating partial entanglement by
  local operations},\ }\href {https://doi.org/10.1103/PhysRevA.53.2046}
  {\bibfield  {journal} {\bibinfo  {journal} {Phys. Rev. A}\ }\textbf {\bibinfo
  {volume} {53}},\ \bibinfo {pages} {2046} (\bibinfo {year}
  {1996})}\BibitemShut {NoStop}%
\bibitem [{\citenamefont {Dada}\ \emph {et~al.}(2011)\citenamefont {Dada},
  \citenamefont {Leach}, \citenamefont {Buller}, \citenamefont {Padgett},\ and\
  \citenamefont {Andersson}}]{dada2011nphys}%
  \BibitemOpen
  \bibfield  {author} {\bibinfo {author} {\bibfnamefont {A.~C.}\ \bibnamefont
  {Dada}}, \bibinfo {author} {\bibfnamefont {J.}~\bibnamefont {Leach}},
  \bibinfo {author} {\bibfnamefont {G.~S.}\ \bibnamefont {Buller}}, \bibinfo
  {author} {\bibfnamefont {M.~J.}\ \bibnamefont {Padgett}},\ and\ \bibinfo
  {author} {\bibfnamefont {E.}~\bibnamefont {Andersson}},\ }\bibfield  {title}
  {\bibinfo {title} {Experimental high-dimensional two-photon entanglement and
  violations of generalized bell inequalities},\ }\href
  {https://doi.org/https://doi.org/10.1038/nphys1996} {\bibfield  {journal}
  {\bibinfo  {journal} {Nature Physics}\ }\textbf {\bibinfo {volume} {7}},\
  \bibinfo {pages} {677} (\bibinfo {year} {2011})}\BibitemShut {NoStop}%
\bibitem [{\citenamefont {Costa}\ \emph {et~al.}(2018)\citenamefont {Costa},
  \citenamefont {Uola},\ and\ \citenamefont {G\"uhne}}]{costa18pra-tsallis}%
  \BibitemOpen
  \bibfield  {author} {\bibinfo {author} {\bibfnamefont {A.~C.~S.}\
  \bibnamefont {Costa}}, \bibinfo {author} {\bibfnamefont {R.}~\bibnamefont
  {Uola}},\ and\ \bibinfo {author} {\bibfnamefont {O.}~\bibnamefont
  {G\"uhne}},\ }\bibfield  {title} {\bibinfo {title} {Steering criteria from
  general entropic uncertainty relations},\ }\href
  {https://doi.org/10.1103/PhysRevA.98.050104} {\bibfield  {journal} {\bibinfo
  {journal} {Phys. Rev. A}\ }\textbf {\bibinfo {volume} {98}},\ \bibinfo
  {pages} {050104} (\bibinfo {year} {2018})}\BibitemShut {NoStop}%
\bibitem [{\citenamefont {Piani}\ and\ \citenamefont
  {Watrous}(2009)}]{piani09prl-channel}%
  \BibitemOpen
  \bibfield  {author} {\bibinfo {author} {\bibfnamefont {M.}~\bibnamefont
  {Piani}}\ and\ \bibinfo {author} {\bibfnamefont {J.}~\bibnamefont
  {Watrous}},\ }\bibfield  {title} {\bibinfo {title} {All entangled states are
  useful for channel discrimination},\ }\href
  {https://doi.org/10.1103/PhysRevLett.102.250501} {\bibfield  {journal}
  {\bibinfo  {journal} {Phys. Rev. Lett.}\ }\textbf {\bibinfo {volume} {102}},\
  \bibinfo {pages} {250501} (\bibinfo {year} {2009})}\BibitemShut {NoStop}%
\bibitem [{\citenamefont {Srivastav}\ \emph {et~al.}(2022)\citenamefont
  {Srivastav}, \citenamefont {Valencia}, \citenamefont {McCutcheon},
  \citenamefont {Leedumrongwatthanakun}, \citenamefont {Designolle},
  \citenamefont {Uola}, \citenamefont {Brunner},\ and\ \citenamefont
  {Malik}}]{22arxiv-malik-noise}%
  \BibitemOpen
  \bibfield  {author} {\bibinfo {author} {\bibfnamefont {V.}~\bibnamefont
  {Srivastav}}, \bibinfo {author} {\bibfnamefont {N.~H.}\ \bibnamefont
  {Valencia}}, \bibinfo {author} {\bibfnamefont {W.}~\bibnamefont
  {McCutcheon}}, \bibinfo {author} {\bibfnamefont {S.}~\bibnamefont
  {Leedumrongwatthanakun}}, \bibinfo {author} {\bibfnamefont {S.}~\bibnamefont
  {Designolle}}, \bibinfo {author} {\bibfnamefont {R.}~\bibnamefont {Uola}},
  \bibinfo {author} {\bibfnamefont {N.}~\bibnamefont {Brunner}},\ and\ \bibinfo
  {author} {\bibfnamefont {M.}~\bibnamefont {Malik}},\ }\bibfield  {title}
  {\bibinfo {title} {Noise-robust and loss-tolerant quantum steering with
  qudits}\ }\href {https://doi.org/arXiv:2202.09294} {arXiv:2202.09294}
  (\bibinfo {year} {2022})\BibitemShut {NoStop}%
\end{thebibliography}%

\end{document}